\documentclass[aps, prl, english, 
reprint, 
showpacs, preprintnumbers,amsmath,amssymb,floatfix,superscriptaddress]{revtex4-1}
\pdfoutput=1

\usepackage[T1]{fontenc}
\usepackage[latin9]{inputenc}
\usepackage{pstricks,pstricks-add} 
\usepackage{graphicx}
\usepackage{amssymb}
\usepackage{babel}
\usepackage{verbatim}
\usepackage[colorlinks=true,urlcolor=blue,linkcolor=red,citecolor=red,breaklinks=true]{hyperref} 
\usepackage{breakurl}
\makeatletter

\usepackage[version=3]{mhchem}

\usepackage{color}

\usepackage{ulem}

\makeatother
\usepackage{babel}
\makeatother

\begin{document}
\title{ Lifshitz interaction can promote ice growth at water-silica interfaces}

\author{Mathias Bostr{\"o}m}
\email{Mathias.A.Bostrom@ntnu.no}
\affiliation{Department of Energy and Process Engineering, Norwegian University of Science and Technology, NO-7491 Trondheim, Norway}
\affiliation{Centre for Materials Science and Nanotechnology, University of Oslo, P. O. Box 1048 Blindern, NO-0316 Oslo, Norway}

\author{Oleksandr I. Malyi}
\affiliation{School of Materials Science and Engineering, Nanyang Technological University, 50 Nanyang Avenue, Singapore 639798, Singapore}

\author{Prachi Parashar}
\email{prachi.parashar@ntnu.no}
\affiliation{Department of Energy and Process Engineering, Norwegian University of Science and Technology, NO-7491 Trondheim, Norway}
\affiliation{Department of Physics, Southern Illinois University-Carbondale, Carbondale, Illinois 62901 USA}

\author{K. V. Shajesh}
\affiliation{Department of Energy and Process Engineering, Norwegian University of Science and Technology, NO-7491 Trondheim, Norway}
\affiliation{Department of Physics, Southern Illinois University-Carbondale, Carbondale, Illinois 62901 USA}

\author{Priyadarshini Thiyam}
\affiliation{Department of Materials Science and Engineering, Royal Institute of Technology, SE-100 44 Stockholm, Sweden}

\author{Kimball A. Milton}
\affiliation{Homer L. Dodge Department of Physics and Astronomy, University of Oklahoma, Norman, Oklahoma 73019, USA}

\author{Clas Persson}
\affiliation{Centre for Materials Science and Nanotechnology, University of Oslo, P. O. Box 1048 Blindern, NO-0316 Oslo, Norway}
\affiliation{Department of Materials Science and Engineering, Royal Institute of Technology, SE-100 44 Stockholm, Sweden}
\affiliation{Department of Physics, University of Oslo, P. O. Box 1048 Blindern, NO-0316 Oslo, Norway}

\author{Drew F. Parsons}
\affiliation{School of Engineering and Information Technology, Murdoch University, 90 South Street, Murdoch, WA 6150, Australia}

\author{Iver Brevik}
\email{iver.h.brevik@ntnu.no}
\affiliation{Department of Energy and Process Engineering, Norwegian University of Science and Technology, NO-7491 Trondheim, Norway}

\vspace{2mm}

\begin{abstract}
At air-water interfaces, the Lifshitz interaction by itself does not 
promote ice growth. On the contrary, we find that the Lifshitz 
force promotes the growth of an ice film, up to 1--8 nm thickness, 
near silica-water interfaces at the triple point of water. This is 
achieved in a system where the combined effect of the retardation 
and the zero frequency mode influences the short-range interactions 
at low temperatures, contrary to common understanding. 
Cancellation between the positive and negative contributions in the 
Lifshitz spectral function is reversed in silica with high porosity.
Our results 
provide a model for how water freezes on glass and other surfaces. 
\end{abstract}

\maketitle

Although water in its different forms has been studied for a very long time, 
several properties of water and ice remain uncertain and are currently under 
intense investigation~\cite{Arbe,Gillan,Benet,Lintunen}.
The question we want to address in the present paper is to what extent the 
fluctuation-induced Lifshitz interaction can promote the growth of ice films 
at water-solid interfaces, at the triple point of water. Particles and surfaces, 
e.g., quartz, soot, or bacteria, in supercooled water are known experimentally 
to nucleate ice formation~\cite{Pandey,Murray,Hoose}. Here, we focus on 
interfaces between water and silica-based materials and examine the roles of 
several intervening factors in the sum over frequency modes (Matsubara terms) 
contributing to the Lifshitz free energy. 

\begin{figure}
\begin{center}
\includegraphics{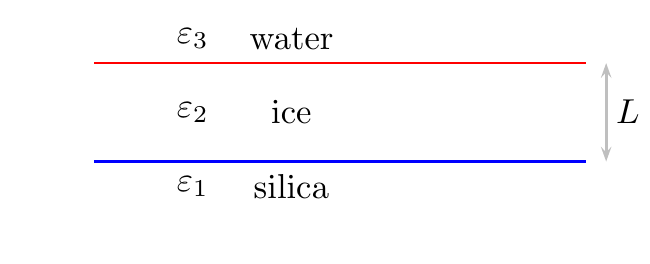}
\caption{Ice ($\varepsilon_2$) of thickness $L$ at the interface of 
water ($\varepsilon_3$) and silica ($\varepsilon_1$),
illustrated here as three planar regions of infinite extent.}
\label{fig-planes-123}
\end{center}
\end{figure}

Quantum fluctuations in the electromagnetic field result in van 
der Waals interactions, which in their unretarded form were explained 
by London in terms of frequency-dependent responses to the fluctuations 
in the polarizable atoms constituting the material medium~\cite{London1,*London2,*Hettema}. 
The understanding of these interactions was revolutionized when Casimir 
introduced retardation effects~\cite{Casi}. The theory was later generalized 
by Lifshitz to include dielectric materials~\cite{Lif,Dzya}. The 
Lifshitz formula in Eq.~(\ref{equ1}), derived for three-layer planar 
geometries~\cite{Dzya}, gives the interaction energy between two semi-infinite 
dielectric media described by their frequency-dependent dielectric permittivities 
as well as the dielectric permittivity of the medium separating them (see Fig.~\ref{fig-planes-123}).
%
%

The purpose of the present work is twofold. First, we want to show 
that a finite size ice film, nucleated by a solid-water interface,
can be energetically favorable even when only the Lifshitz interaction 
is accounted for. Second, we want to highlight a relevant contribution 
from the zero frequency term in the expression for the Lifshitz energy 
in a region where it is not expected to be important. The temperature 
dependence of the Casimir force between metal surfaces~\cite{Dzya,Bost2000,Bord,Hoye} 
relies strongly on the exact behavior of the low-frequency dielectric 
function of metals. These and many other investigations have provided 
support for the notion that the zero frequency term would only be 
relevant at high temperatures or large surface separations at a moderate 
temperature. In biological systems that involve water, the zero frequency 
term contributes substantially to the total Lifshitz interaction energy 
because of the high static dielectric permittivity of the water compared 
to the interacting media~\cite{Isra,Pars1970}.

In this paper, we will show that for three-layer planar geometries, 
where an attractive-repulsive force transition can occur, it is 
possible to find systems in which the combined effect of retardation 
and the zero frequency term determines what happens with the interaction  
across extremely thin sheets. The mechanism behind this is a cancellation 
between the positive (repulsive) and negative (attractive) contributions 
from the different frequency regions, which leads to a diminished 
contribution from the nonzero Matsubara terms and thus renders the zero 
frequency Matsubara term dominant.

We emphasize that the system, in spite of its apparent simplicity, 
is far from trivial. The resulting value of the Lifshitz energy is 
dependent on an interplay between different factors:

(i) The crossing in the curves for the permittivities $\varepsilon$ 
as functions of the imaginary frequency $\zeta$, where the crossing occurs in 
the optical 
region, results in a switch from attractive to repulsive contributions to the 
Lifshitz force.

(ii) The need to include retardation effects in the formalism: This 
may appear surprising, as retardation effects due to the finite speed of light 
$c$ are usually related to cases where the gap widths are large.

(iii) The dominant role played by the zero frequency Matsubara 
term $n=0$ , which is a direct consequence of the aforementioned two 
factors: This may also be somewhat unexpected, in view of the circumstance 
that the $n=0$ term is usually taken to be important only in the limits of 
large separation distance $L$ at a moderate temperature $T$, or high temperature 
at moderate separation. (Observe that in the special case of a nondispersive 
medium the single nondimensional parameter of importance in the Lifshitz 
sum-integral is $Lk_BT/\hbar c$.)

\begin{figure}
\begin{center}
\includegraphics[width=7.6cm]{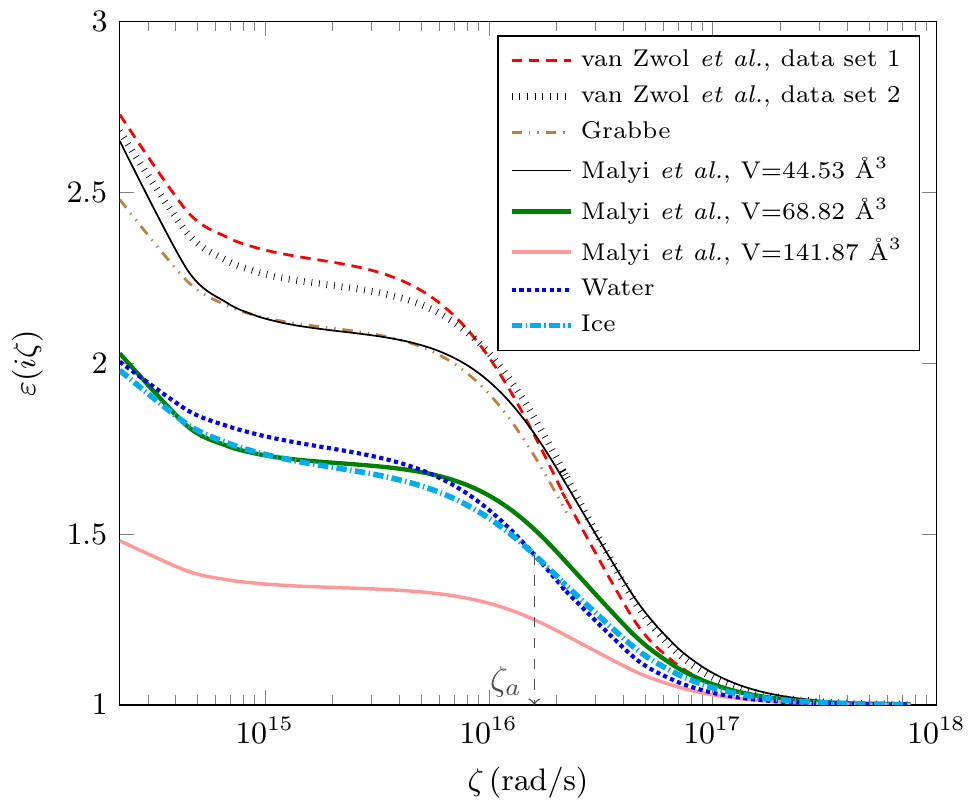}
\caption{(Color online) Permittivity as a function of frequency for 
ice, water, and different silica materials. 
The static values $\varepsilon(0)$ for ice and water are $91.5$ and  $88.2$, 
respectively, using data from Elbaum and Schick~\cite{Elbaum}.  For different 
SiO$_2$ materials, the static values are $3.90$, $2.62$, and $1.69$ using  data from 
Malyi {\it et al.}~\cite{Sasha} for volumes $44.53$, $68.82$, and 
$141.87$ S\AA$^3$, respectively (here extended to include phonon contributions), 
$3.80$ from Grabbe~\cite{Grabbe}, and $3.90$ from data set 1 and data set 2 of 
van Zwol and Palasantzas~\cite{Zwol1}. The transition frequency, 
$\zeta_a\approx1.60\times10^{16}$ rad/s, is where the permittivities of ice and water 
cross in the optical frequency region.
}
\label{figu1}
\end{center}
\end{figure}

The need to include all these effects stems of course from the complicated 
Lifshitz sum-integral, when the dispersive properties of the material components 
are accounted for accurately. In a three-layer planar system, where medium $1$ is interacting with medium $3$ across medium $2$, the system tries to minimize the interaction 
energy, which manifests as a force of attraction if 
$[\varepsilon_1(i\zeta) - \varepsilon_2(i\zeta)]
[\varepsilon_3(i\zeta) - \varepsilon_2(i\zeta)] >0$ 
and a force of repulsion for 
$[\varepsilon_1(i\zeta) - \varepsilon_2(i\zeta)]
[\varepsilon_3(i\zeta) - \varepsilon_2(i\zeta)] <0$. These conditions for attraction and repulsion must hold over a wide 
frequency range because they occur within the Lifshitz sum-integral. The 
plausibility of the repulsive Lifshitz force between two dielectric 
objects with an intervening medium of suitable dielectric permittivity 
was first discussed by Dzyaloshinskii {\it et al.}~\cite{Dzya} and has 
been observed experimentally~\cite{AndSab,Mun,Milling,Lee,Feiler}. Earlier 
experimental and theoretical studies are comprehensively discussed 
in Ref.~\cite{Mahanty}. Elbaum and Schick observed that the difference 
between the dielectric permittivities of ice and water changes sign at 
the transition frequency ($\zeta_a\,\approx1.60\times10^{16}$ rad/s), as shown 
in Fig.~\ref{figu1}~\cite{Elbaum}. 
%
Thus, the contribution to the Lifshitz force, above and below the 
transition frequency $\zeta_a$, is attractive and repulsive in nature, 
respectively. Furthermore, the difference between the dielectric 
permittivities of ice and water changes sign again at frequencies lower 
than the first Matsubara frequency, thus affecting the overall behavior 
of the Lifshitz force. Elbaum and Schick showed that these attractive 
and repulsive contributions for the ice-water-vapor system, at the 
triple point, lead to the formation of a thin layer of water at the 
interface of ice and vapor~\cite{Elbaum}. The scale for the thickness 
of the layer of water is set by the transition distance $c/\zeta_a$. 
Most often it is argued that the retardation effects can be neglected 
if the distance is less than a few tens of nanometers. However, several 
studies~\cite{Rich1,bosserPRA2012,Elbaum,Zwol1,Ninh} highlight the 
importance of including the retardation effect even at the separation 
distances of less than $10$ nm.

We investigate if a thin layer of ice at the interface of silica and water 
will grow (freeze) or vanish (melt), near the triple point of water, assisted 
exclusively by the Lifshitz interaction. In Ref.~\cite{Elbaum2}, Elbaum and Schick 
find that a thin sheet of ice does not grow at the water-vapor interface. 
In contrast, we report that the Lifshitz force does assist ice growth at the 
silica-water interface. The thickness of the ice layer formed at the silica-water 
interface varies with the permittivities of the silica substrate. 
(In Ref.~\cite{Dash2006}, Dash {\it et al.} thoroughly reviewed a related phenomenon 
of the premelting of ice, which was also considered by some of us in Ref.~\cite{MathiasEPL2016} 
where we showed that it is essential to have a vapor layer between ice and a silica 
surface to have premelting of the ice.)

To study ice growth at the silica-water interface, we consider a model system 
with a planar silica surface interacting with water across a thin planar 
ice film of thickness $L$, as illustrated in Fig.~\ref{fig-planes-123}. 
The ice sheet thicknesses that we discuss are typically in the range 1--8 nm. 
Recently, Schlaich {\it et al.}~\cite{Netz} showed that the dielectric functions 
for films thicker than 1 nm approached their bulk values. Thus, to predict 
trends, it should be sufficient to use bulk dielectric functions for the 
thin ice layer. The Lifshitz interaction free energy per unit area $F$ is expressed 
as a sum of Matsubara frequencies,  $\zeta_n = 2\pi n/\hbar\beta$~\cite{Dzya},
\begin{equation}
F(L)={\sum_{n = 0}^\infty}^\prime g(L,i\zeta_n), \quad \beta =\frac{1}{k_BT},
\label{equ1}
\end{equation}
where $g(L,i\zeta_n)$ obtains contributions from the transverse electric (TE) 
and the transverse magnetic (TM) modes,
\begin{eqnarray}
g(L,i\zeta_n) &=&\frac{1}{\beta} \int\frac{d^2k}{(2\pi)^2}
\big\{\ln
\left[1 - e^{-2\gamma_2 L} r_{21}^\text{TE} r_{23}^\text{TE} \right]
\nonumber \\ && +
\ln\left[1 - e^{-2\gamma_2 L} r_{21}^\text{TM} r_{23}^\text{TM} \right]\big\}.
\label{equ2}
\end{eqnarray}
Here, $\gamma _i= \sqrt{k^2 +{{\left( {\zeta_n/c} \right)}^2}{\varepsilon _i}}$, 
$k$ is the magnitude of the wave vector parallel to the surface, and the prime 
on the summation sign indicates that the $n=0$ term should be divided by 2.
We have used the notations
\begin{equation}
r_{ij}^\text{TE} = \frac{\gamma _i - \gamma _j}{\gamma _i + \gamma _j}
\quad \text{and} \quad
r_{ij}^\text{TM}=\frac{{{\varepsilon _j}{\gamma _i} - {\varepsilon _i}{\gamma _j}}}{{{\varepsilon _j}{\gamma _i} + {\varepsilon _i}{\gamma _j}}}
\label{equ3}
\end{equation}
for the TE and TM mode reflection coefficients.

We use dielectric functions for different silica,  each with a specific 
nanoporosity, or average volume ($V$) per SiO$_2$ unit, computed directly 
from first-principles calculations, as reported in our previous work~\cite{Sasha}. 
However, since the phonon contribution to the 
dielectric function at imaginary frequencies can have a noticeable 
impact on the Lifshitz forces, we model the phonon parts of the 
dielectric functions using the single-phonon Lorentz model and the 
Kramers-Heisenberg equation~\cite{Klingshirn}. Here, the longitudinal 
frequency ($\zeta_\text{LO}=0.1351$ eV), taken to be the same for all considered 
systems, is determined from the fitting of the multiphonon contribution 
to the dielectric function of quartz. The longitudinal and transverse 
optical frequencies for quartz are taken directly from the experimental 
data~\cite{quartz}.  At the same time, the single-phonon transverse 
frequency $\zeta_\text{TO}$ is computed from 
the Lyddane-Sachs-Teller equation~\cite{Lydanne} using the fitted 
$\zeta_\text{LO}$ and dielectric constants reported in our previous 
work~\cite{Sasha}. We also use parametrized model 
dielectric functions for different silica materials based on the 
optical data and the Kramers-Kronig relationship given 
by Grabbe~\cite{Grabbe} and two separate data sets by van Zwol and 
Palasantzas~\cite{Zwol1} for comparison. We take the dielectric 
functions of ice and water at $T=273.16$ K from Elbaum and 
Schick~\cite{Elbaum}. Figure~\ref{figu1} shows the plots of 
dielectric functions for ice, water, and different silica materials.
\begin{figure}
\begin{center}
\includegraphics[width=7.6cm]{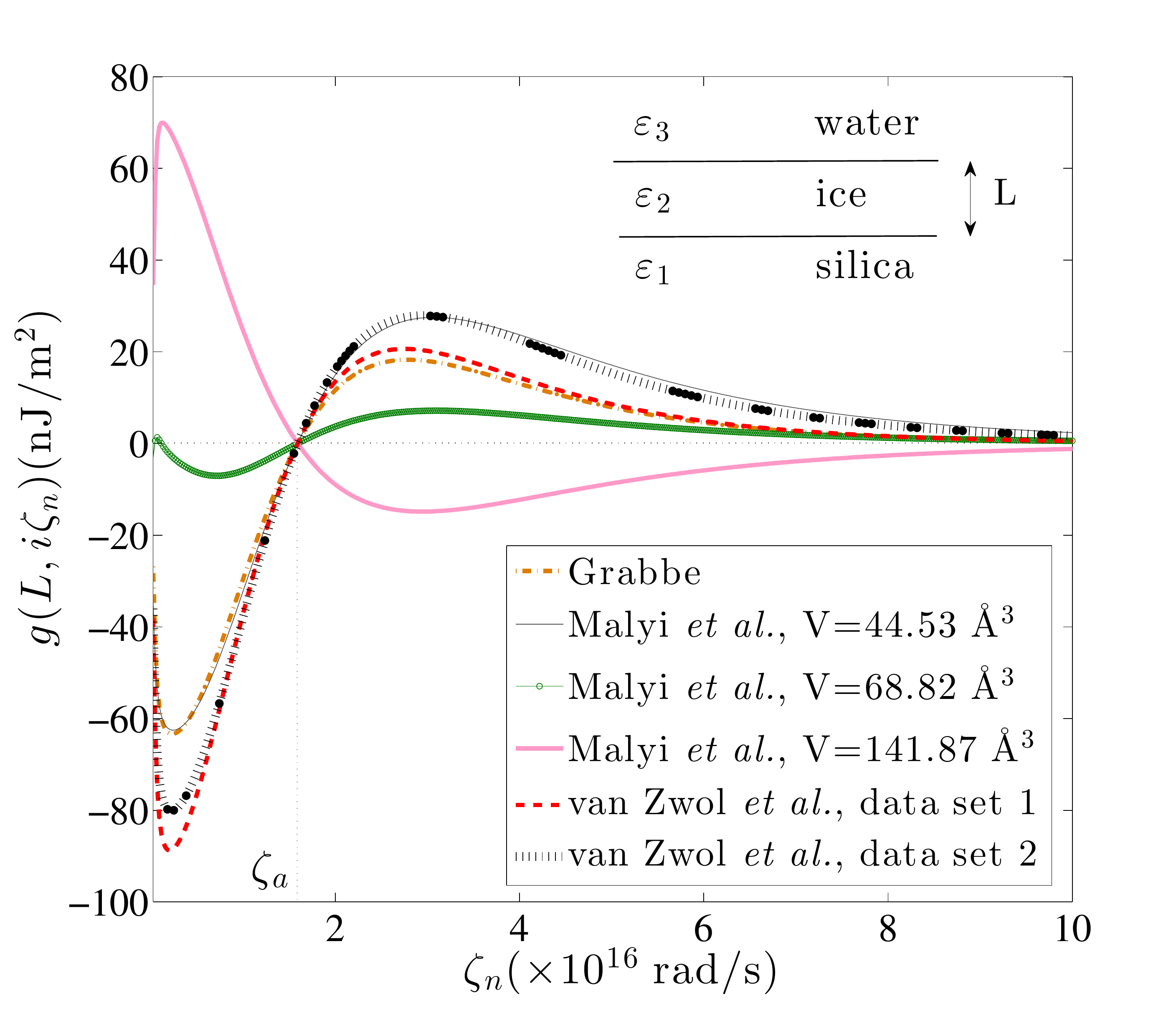}
\caption{(Color online) Spectral function $g(L,i\zeta_n)$ as a function of Matsubara
frequency ($\zeta_n$) for a silica-ice-water system with $L=2.0$ nm thick ice 
film. We compare the result using different silica dielectric functions presented 
in Fig.~\ref{figu1}. The zero frequency contributions for various silica materials 
(not shown in the figure) are at least one order of magnitude higher than the 
contributions from other Matsubara frequencies ($\approx-300$ nJ/m$^2$). All the 
curves vanish at the same point corresponding to the transition frequency  
$\zeta_a\approx1.60\times10^{16}$ rad/s.}
\label{figu2}
\end{center}
\end{figure}

In Fig.~\ref{figu2} we plot the spectral function $g(L,i\zeta_n)$ in 
Eq.~(\ref{equ2}), for different silica-ice-water systems, at $L=2.0$ nm. 
The total Lifshitz energy is the area under the curve(s), getting positive 
contributions from the positive area and negative contributions from the 
negative area. The cancellation between these contributions results in 
a dominant role for the $n=0$ Matsubara term. In the symmetric systems 
involving water ($\varepsilon_1=\varepsilon_3$) the large static dielectric permittivity 
of water compared to the interacting media causes an increase in the 
factor $r_{21}^\text{TM}(0) r_{23}^\text{TM}(0)\approx0.9$. This enhances significantly the contribution of the $n=0$ term to the total interaction energy~\cite{Isra}. By contrast, in our asymmetric silica-ice-water system the above factor is approximately $0.02$ due to very similar values of the static dielectric permittivities of ice and water. The contribution of the $n=0$ term is therefore not enhanced here.

\begin{figure}
\begin{center}
\includegraphics[width=7.6cm]{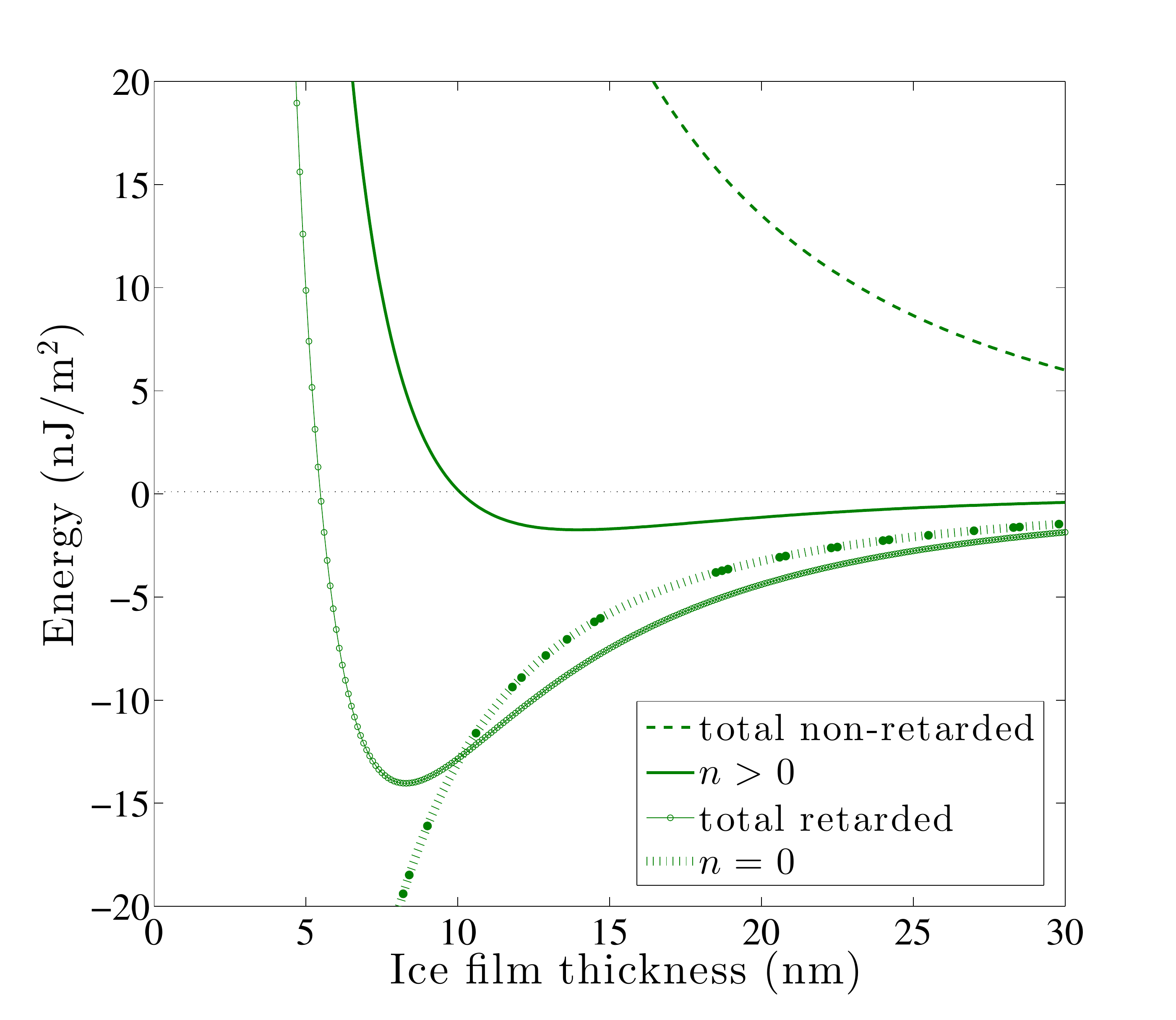}
\caption{(Color online) Contributions to the Lifshitz free energy per 
unit area for a $V=68.82$ \AA$^3$) system as a function of 
ice film thickness. Four different curves are shown: the total nonretarded 
energy, the contributions from the $n>0$ term to the retarded energy, the total retarded energy, and 
the contribution from the $n=0$ term alone. }
\label{figu3}  
\end{center}
\end{figure}

We nevertheless find that the $n=0$ Matsubara term is 
crucial for all separation distances, as shown in Fig.~\ref{figu3}. It is 
evident from the plot that if we ignore the retardation effect, then there 
will be a complete freezing of the water, which, however, is not a natural 
phenomenon. The contribution to the Lifshitz energy from the $n=0$ term is 
always attractive and considerably influences the equilibrium thickness as 
well as the stability of the ice sheet as compared to the contributions from 
the $n>0$. This conclusion is true for most materials with a low dielectric 
constant that can serve as nucleation sites for ice formation but not 
for metals, where the $n=0$ term gives a repulsive contribution.

\begin{figure}
\begin{center}
\includegraphics[width=8cm]{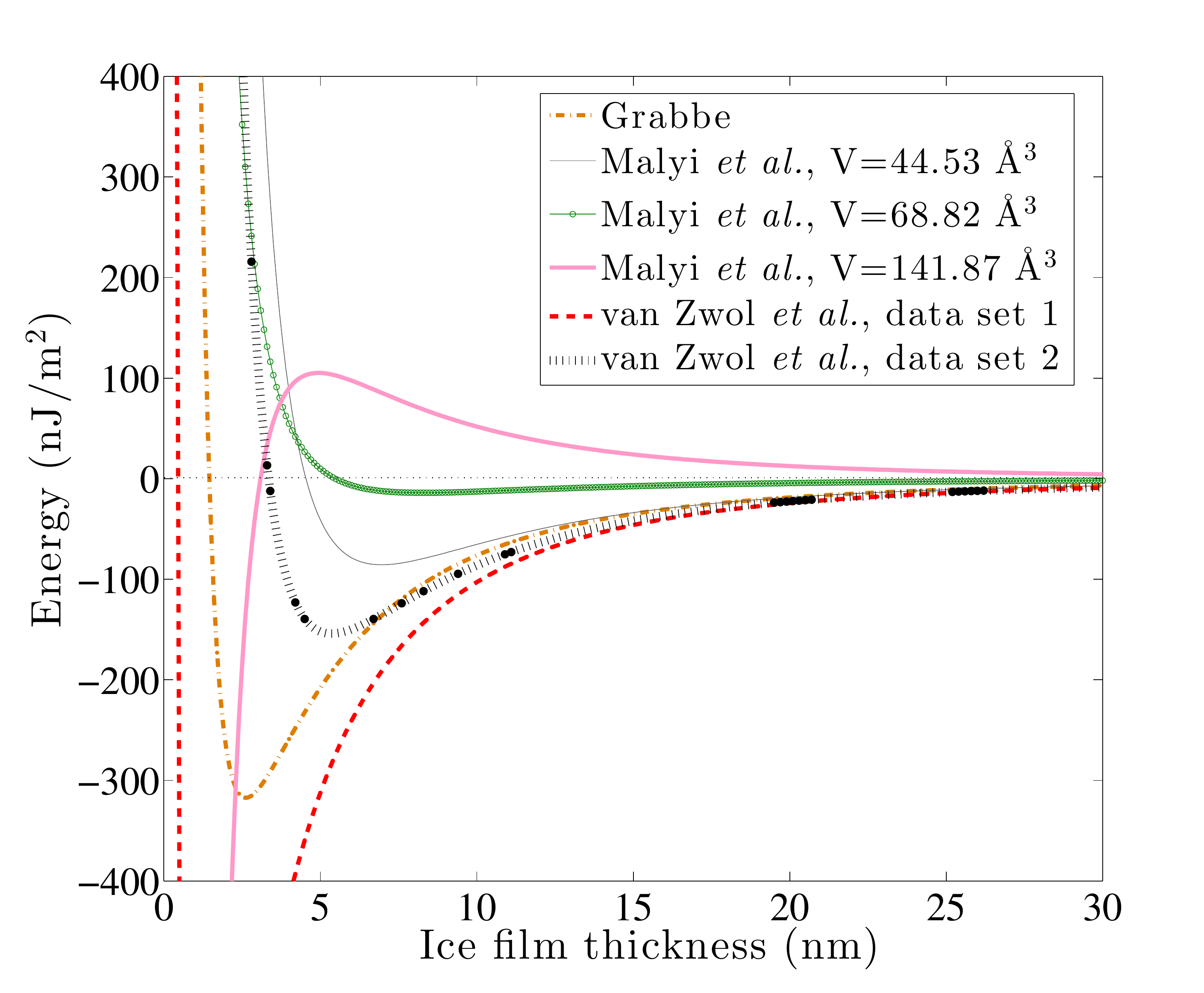}
\caption{(Color online) Lifshitz free energy per unit area for silica-ice-water 
systems as a function of the ice film thickness using different silica dielectric functions presented 
in Fig.~\ref{figu1}. }
\label{figu4}
\end{center}
\end{figure}

An estimate for the stable thickness of ice formed at the interface of the 
silica-water system is obtained~\cite{Ninh} by replacing the two exponentials 
in Eq.~(\ref{equ2}) with a step function, $e^x\sim\theta(x)$. This corresponds 
to $2\gamma_2 L\approx 1$, which leads to 
$L\approx c/2 \zeta_a \sqrt{\varepsilon_2(\zeta_a)}= 7.9\,$nm. This 
estimate is similar to the equilibrium thicknesses of the ice sheets 
for the broad range of the silica-water interfaces calculated using the complete 
Lifshitz energy of Eq.~(\ref{equ1}), shown in Table~\ref{ice_film_thickness}.  
This stable thickness corresponds to an extremum in the plots of the total 
Lifshitz energy versus the separation distance $L$ in Fig.~\ref{figu4}. The 
last column in Table~\ref{ice_film_thickness} shows the relative contribution 
of the $n=0$ term with respect to the total energy at the equilibrium thickness. 
It is clear that the contribution from the $n=0$ term is dominant in most 
cases, even at the small separation distances, and even exceeds the contribution 
coming from the $n>0$ terms in some cases.

Typically for the Casimir interaction between two atoms, retardation effects 
become relevant for distance regimes set by the cube root of the polarizability 
of the atoms, which serves as the scale for the retardation effects. In our 
system, the characteristic frequency is the transition frequency $\zeta_a$, 
which sets the scale for retardation to be $8$ nm. This includes the speed 
of light in the intermediate medium.

\begin{table}
\begin{tabular}{|c|c|c|}
\hline
Volume  (\AA$^3$)& Ice film thickness (nm)  & $F_\text{n=0}^\text{eq}/F^\text{eq}$\\ \hline
35.68 & 7.0 & 0.20 \\
44.53&  7.0 & 0.30\\
56.96 & 7.2 & 0.57 \\
68.82 & 8.3  &  1.34\\
106.39 & 0 & - \\
141.87 &  0  & -\\ \hline
Grabbe  &  2.6  & 0.59\\
data set 1 & 0.9  & 1.29 \\
data set 2 & 5.4  & 0.28\\
\hline
\end{tabular}
\caption{Ice film thickness at different silica-water interfaces at the triple point 
of water. The stable equilibrium ice film 
thickness is shown in the middle column. A zero value corresponds to the absence 
of a stable equilibrium at small distances for high nanoporosity silica material. 
In the last column we show the ratio between the $n=0$ term and the total retarded 
Lifshitz energy at the equilibrium ice film thickness. The plots for volumes 
35.68 and 106.39 \AA$^3$ are not shown in Figs. \ref{figu2} and \ref{figu4}.}
\label{ice_film_thickness}
\end{table}

We summarize our results for ice formation near silica surfaces in Fig.~\ref{figu4} 
and Table~\ref{ice_film_thickness}. We find that the system shows the behavior of 
the vapor-ice-water interface of Ref.~\cite{Elbaum2}, i.e., the intermediate layer 
vanishes, for very high nanoporosity (large $V$ for the SiO$_2$ material). The spectral 
function $g(L,i\zeta_n)$ in this case is reversed (see Fig.~\ref{figu2}). In this 
limit when the substrate behaves more as a vapor, there is no ice growth due to 
Lifshitz forces, as predicted by Elbaum and Schick~\cite{Elbaum2}. For these cases, 
due to the attractive $n=0$ contribution, there is a global energy maximum around 
{$L$=4-5\,nm} and a local very weak energy minimum around {$L$=1--2$ \mu$m}. 
However, for a large range of different silica materials, we predict a surface 
specific ice growth near the silica-water interface. The transition point between a 
stabilized thin ice layer and destabilized ice growth is apparent from the 
dielectric spectrum of nanoporous silica, seen in Fig.~\ref{figu1}. The stable 
thin layer is lost when the silica porosity is high enough to cause its dielectric 
function to remain below that of ice and water. 

The study of ice formation at a silica interface has significant applied value 
as the model system for how water freezes on glass, rocks, and soil surfaces. 
Quasiliquid layers are observed to form on solid-ice interfaces, depending 
on the surface density and roughness~\cite{Engemann,Beaglehole,Andersson,Wilen,Liljeblad}. 
Optical reflection measurements have demonstrated the existence of up to a few tens of 
nanometer thick  premelted water sheets on ice crystal surfaces~\cite{Elbaum3,Dosch,Dash1995,Veen}. 
Ice in contact with silica has been found to have a {5--6 nm} thick quasiliquid layer 
on the surface with a density similar to high-density amorphous ice~\cite{Engemann}. 
Several measurements have been carried out aiming at an understanding of the structure 
of the ice surface~\cite{Slaughterbeck,Pittenger,Petrenko,Doppenschmidt}. For a thorough 
review on the premelting of the ice, see Dash {\it et al.} in Ref.~\cite{Dash2006}. 
From our study, we find that the Lifshitz force promotes freezing in the limit of low porosity, analogous to the reduction 
in the premelting layer observed with decreasing temperature~\cite{Liljeblad}.
In another experimental study, Bluhm and 
Salmeron~\cite{Bluhm} observe a {0.7 nm} thin sheet of ice formed at the mica-water interface. We obtain a thickness {2.7 nm} for ice formation on mica using the above techniques with the dielectric permittivity of mica from Ref.~\cite{Chan}. 

In real systems, optical properties, surface charges, surface 
roughness~\cite{Benet}, the density of the material, 
gravity~\cite{Rich1,EstesoJCP}, ions~\cite{Wilen,Bost2001,Thiyam}, the presence of gas layers on ice premelting in pores~\cite{MathiasEPL2016} and so on 
influence the total energy of the system. It is an advantage of the theory 
that different properties can be analyzed separately. 

In summary, the investigations of ice growth, due to the Lifshitz interaction, near 
different materials require a detailed knowledge of the dielectric functions for 
a large range of frequencies. The zero frequency term, although of fundamental 
interest in its own right, can under specific circumstances also play a major 
role in determining the stability and thickness of a thin layer near surfaces 
at much shorter distances than one would normally expect. Elbaum and Schick 
observed that the Lifshitz interaction is not sufficient, by itself, to promote 
ice growth at the water-vapor surface~\cite{Elbaum2}. In contrast, we predict a 
growth of nanosized ice films driven by the Lifshitz interaction at certain 
silica interfaces in ice-cold water. We suggest that it should be possible to 
measure them, perhaps with the use of already available experimental 
techniques~\cite{Yimin}.

We acknowledge support from the Research Council of Norway (Projects 221469 and 250346).
  We also acknowledge access to high-performance computing resources via SNIC and NOTUR.

\end{document}